%% file: ll2010_arXiv.tex
\documentclass[fleqn,twoside]{article}
\usepackage{amsmath,amssymb,espcrc2,color,cite}

\usepackage{graphicx}
\usepackage[figuresright]{rotating}

\input paperdef
\graphicspath{{figs/}}

\newcommand{\gmt}{$(g-2)_\mu$}
\newcommand{\bsg}{BR($b \to s \gamma$)}
\newcommand{\btn}{BR($B_u \to \tau \nu_\tau$)}
\newcommand{\bmm}{BR($B_s \to \mu^+\mu^-$)}
\newcommand{\ssi}{\sigma^{\rm SI}_p}
\newcommand{\Och}{\ensuremath{\Omega_\chi h^2}}
\newcommand{\ecm}{\sqrt{s}}

\begin{document}

\include{ll2010_titlepage}

\input{ll2010_main}


\end{document}

%% file: ll2010_titlepage.tex
\thispagestyle{empty}
\setcounter{page}{0}
\def\thefootnote{\fnsymbol{footnote}}

\begin{flushright}
\mbox{}
DESY 10-099\\
arXiv:1007.0206 [hep-ph]
\end{flushright}

\vspace{1cm}

\begin{center}

{\large\sc {\bf Predicting Supersymmetry}}
\footnote{invited talk given by S.H.\ at the {\em Loops \& Legs 2010}, 
April 2010, W\"orlitz, Germany}

\vspace{1cm}

{\sc 
S.~Heinemeyer$^{1}$%
\footnote{
email: Sven.Heinemeyer@cern.ch}%
~and G.~Weiglein$^{2}$%
\footnote{email: Georg.Weiglein@desy.de}
}

\vspace*{1cm}

{\it
$^1$Instituto de F\'isica de Cantabria (CSIC-UC), 
Santander,  Spain\\

\vspace{0.3cm}

$^2$DESY, Notkestra\ss e 85, D--22603 Hamburg, Germany
}
\end{center}

\vspace*{0.2cm}

\BC {\bf Abstract} \EC
\input{ll2010_abstract}

\def\thefootnote{\arabic{footnote}}
\setcounter{footnote}{0}

\newpage


%% file: ll2010_abstract.tex
We review the result of SUSY parameter fits based on 
frequentist analyses of experimental constraints
from electroweak precision data, \gmt, $B$ physics and cosmological data.
We investigate the parameters of the constrained MSSM (CMSSM) 
with universal soft supersymmetry-breaking mass parameters, and a model
with common non-universal Higgs mass parameters in the superpotential (NUHM1).
Shown are the results for the SUSY and Higgs spectrum of the models. 
Many sparticle masses are highly correlated in both the CMSSM and NUHM1, 
and parts of the regions preferred at the 68\% C.L.\ are accessible to
early LHC running. The best-fit points could be tested even with 1~\ifb\
at $\sqrt{s} = 7 \tev$.

%% file: ll2010_main.tex
\title{Predicting Supersymmetry}

\author{S.~Heinemeyer\address[ifca]
          {Instituto de F\'{\i}sica de
          Cantabria (CSIC-UC), E--39005 Santander, Spain},
        G.~Weiglein\address[desy]
          {DESY, Notkestrasse 85, D--22603 Hamburg, Germany}
}
       
\begin{abstract}
\input{ll2010_abstract}   
\vspace{1pc}
\end{abstract}

\maketitle


\section{INTRODUCTION}

Supersymmetry (SUSY) is one of the favored ideas for physics beyond
the Standard Model (SM) that may soon be explored at the Large Hadron
Collider (LHC). Here we review the results from frequentist
analyses~\cite{Master3,Master2} of 
the parameter spaces of the constrained minimal supersymmetric
extension of the Standard Model (CMSSM) --- in which the
soft SUSY-breaking scalar and gaugino masses are each constrained
to universal values $m_0$ and $m_{1/2}$, respectively 
 --- and the NUHM1 --- in which the soft SUSY-breaking
contributions to the Higgs masses are allowed to have a different but
common value. Both models have a common trilinear coupling $A_0$ at the
GUT scale and $\tb$ (the ratio of the two vacuum expectation values) as
a low-energy input. 
An overview about the existing literature on this subject can be found
in \citere{Master3}. 

The fit includes precision electroweak data,
the anomalous magnetic moment of the muon, \gmt, 
$B$-physics observables (such as rates for \bsg\ and \btn,
and the upper limit on \bmm), $K$-physics observables, the bound on the
lightest MSSM Higgs boson mass, $\Mh$, and the cold dark matter (CDM) density
inferred from astrophysical and cosmological data%
\footnote{We did not 
include the constraint imposed by the experimental
upper limit on the spin-independent DM scattering
cross section $\ssi$, which is subject to astrophysical and hadronic
uncertainties.}%
, assuming that this is
dominated by the relic density of the lightest neutralino, $\Och$.


\section{THE MASTERCODE}

We define a global $\chi^2$ likelihood function, which combines all
theoretical predictions with experimental constraints:
\begin{align}
\chi^2 &= \sum^N_i \frac{(C_i - P_i)^2}{\sigma(C_i)^2 + \sigma(P_i)^2}
\nonumber \\[.2em]
&+ {\chi^2(\Mh) + \chi^2(\br(B_s \to \mu\mu))}
\nonumber \\[.2em]
&+ {\chi^2(\mbox{SUSY search limits})}
\nonumber \\[.2em]
&+ \sum^M_i \frac{(f^{\rm obs}_{{\rm SM}_i}
              - f^{\rm fit}_{{\rm SM}_i})^2}{\sigma(f_{{\rm SM}_i})^2}~.
\label{eqn:chi2}
\end{align} 
Here $N$ is the number of observables studied, $C_i$ represents an
experimentally measured value (constraint), and each $P_i$ defines a
prediction for the corresponding constraint that depends on the
supersymmetric parameters.
The experimental uncertainty, $\sigma(C_i)$, of each measurement is
taken to be both statistically and systematically independent of the
corresponding theoretical uncertainty, $\sigma(P_i)$, in its
prediction. We denote by
$\chi^2(\Mh)$ (see below) and $\chi^2(\br(B_s \to \mu\mu))$ the $\chi^2$
contributions from two measurements for which only one-sided
bounds are available so far. Similarly, 
we include the lower limits from the direct searches
for SUSY particles at LEP~\cite{LEPSUSY} as one-sided limits, denoted by 
``$\chi^2(\mbox{SUSY search limits})$'' in \refeq{eqn:chi2}.

In order to derive a reliable theoretical prediction of an observable
(denoted as $P_i$ in \refeq{eqn:chi2}) and to obtain a correspondingly
small theory (intrinsic) uncertainty (denoted as $\si(P_i)$ in
\refeq{eqn:chi2}) it 
is crucial to have higher-order corrections (beyond the one-loop level)
at hand. As an example, including SM corrections up to four-loop and
MSSM corrections up to the two-loop level, the theory uncertainties of
the $W$~boson mass, $\MW$, and the effective leptonic weak mixing angle,
$\sweff$ are now at the level of (depending in the MSSM on the SUSY mass scale)
\begin{align}
\de\MW^{\rm intr,SM} &= 4 \mev \mbox{\cite{deMWSM}}~, \\
\de\MW^{\rm intr,MSSM} &= 5-10 \mev \mbox{\cite{MWMSSM}}~, \\
\de\sweff^{\rm intr,SM} &= 4.7 \times 10^{-5} \mbox{\cite{desweffSM}}~,\\
\de\sweff^{\rm intr,MSSM} &= (4.9 - 7.1) \times 10^{-5} 
                                    \mbox{\cite{ZObsMSSM}}~.
\end{align}
Other examples are \bsg, where in the SM the intrinsic uncertainty is
now at the level of $\sim 0.27 \times 10^{-4}$~\cite{debsg} (adding
errors in quadrature) , or \gmt\
with a theory uncertainty of less than $4.9 \times 10^{-10}$~\cite{degm2}. 
Many calculations and tools used in our analyses were presented at 
``Loops \& Legs'' conferences over the past decade, see
\citere{LL2000-2008}. 

The main computer code for our evaluations is the 
{\tt MasterCode}~\cite{Master1,Master2,Master3,Master35,MasterWWW}, 
which includes the following theoretical codes. For the RGE running of
the soft SUSY-breaking parameters, it uses
{\tt SoftSUSY}~\cite{Allanach:2001kg}, which is combined consistently
with the codes used for the various low-energy observables:
{\tt FeynHiggs}~\cite{Heinemeyer:1998yj,Heinemeyer:1998np,Degrassi:2002fi,Frank:2006yh}  
is used for the evaluation of the Higgs masses and  
$a_\mu^{\rm SUSY}$  (see also
\cite{Moroi:1995yh,PomssmRep}),
for the other electroweak precision data we have included 
a code based on~\cite{MWMSSM,ZObsMSSM},
{\tt SuFla}~\cite{Isidori:2006pk,Isidori:2007jw} and 
{\tt SuperIso}~\cite{Mahmoudi:2008tp,Eriksson:2008cx}
are used for flavor-related observables, 
and for dark-matter-related observables
{\tt MicrOMEGAs}~\cite{Belanger:2006is} and
{\tt DarkSUSY}~\cite{Gondolo:2005we} are included.
In the combination of the various codes,
{\tt MasterCode} makes extensive use of the SUSY
Les Houches Accord~\cite{Skands:2003cj,Allanach:2008qq}.

The experimental constraints used in our analyses are listed in
Table~1 of \citere{Master3}. 
One important comment concerns our implementation of the LEP
constraint on $\Mh$. The value quoted in the Table~1 of \citere{Master3}, 
$\MH > 114.4 \gev$,
was derived within the SM~\cite{Barate:2003sz}, and is applicable to the
CMSSM, in which 
the relevant Higgs couplings are very similar to those in the 
SM~\cite{Ellis:2001qv,Ambrosanio:2001xb}, so that the SM
exclusion results can be used, supplemented with an additional theoretical
uncertainty:
we evaluate the $\chi^2(\Mh)$ contribution within the CMSSM using the
formula
\begin{align}
\chi^2(\Mh) = \frac{(\Mh - \Mh^{\rm limit})^2}{(1.1 \gev)^2 + (1.5 \gev)^2}~,
\label{chi2Mh}
\end{align}
with $\Mh^{\rm limit} = 115.0 \gev$ 
for $\Mh < 115.0 \gev$. 
Larger masses do not receive a $\chi^2(\Mh)$ contribution. 
We use $115.0 \gev$ so as to take into account
experimental systematic effects.
The $1.5 \gev$ in the denominator corresponds to a convolution of
 the likelihood function with a Gaussian function, $\tilde\Phi_{1.5}(x)$,
normalized to unity and centered around $\Mh$, whose width is $1.5 \gev$,
representing the theory uncertainty on $\Mh$~\cite{Degrassi:2002fi}.
In this way, a theoretical uncertainty of up to $3 \gev$ is assigned for 
$\sim 95\%$ of all $\Mh$ values corresponding to one CMSSM parameter point. 
The $1.1 \gev$ term in the denominator corresponds to a parameterization
of the $CL_s$ curve given in the final SM LEP Higgs
result~\cite{Barate:2003sz}. 
Within the NUHM1 the situation is somewhat more involved, since, for
instance, a strong suppression of the $ZZh$ coupling can occur,
invalidating the SM exclusion bounds. 
In order to find a more reliable 95\% C.L.\ exclusion limit for $\Mh$ in the
case that the SM limit cannot be applied, we use a specially adopted procedure
outlined in \citere{Master3} (using the results of the code {\tt
HiggsBounds}~\cite{higgsbounds} that incorporates the
LEP~\cite{Schael:2006cr} (and Tevatron) 
limits on neutral Higgs boson searches).

\begin{figure*}[htb!]
\begin{center}
\includegraphics[width=0.48\textwidth]{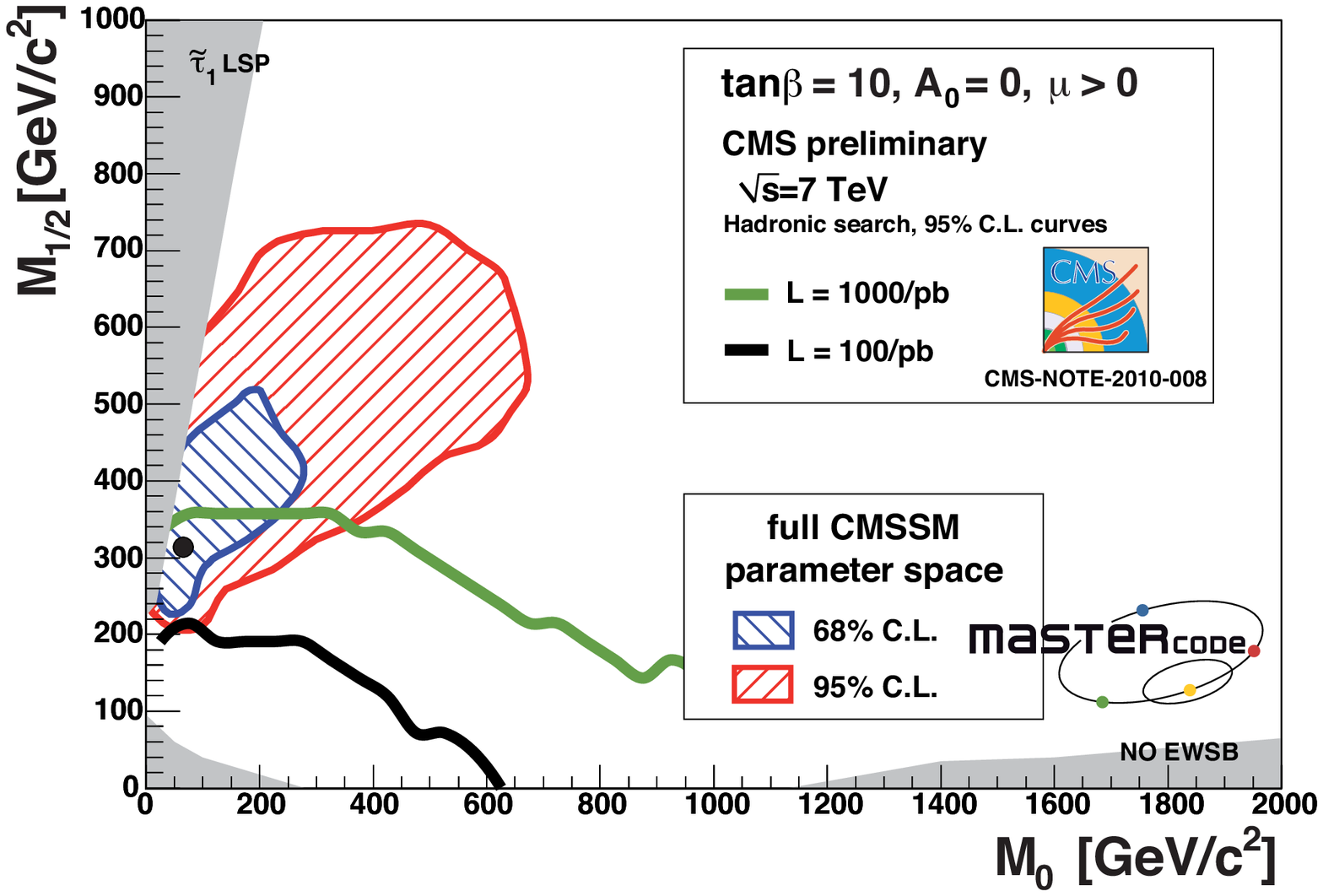}
\includegraphics[width=0.48\textwidth]{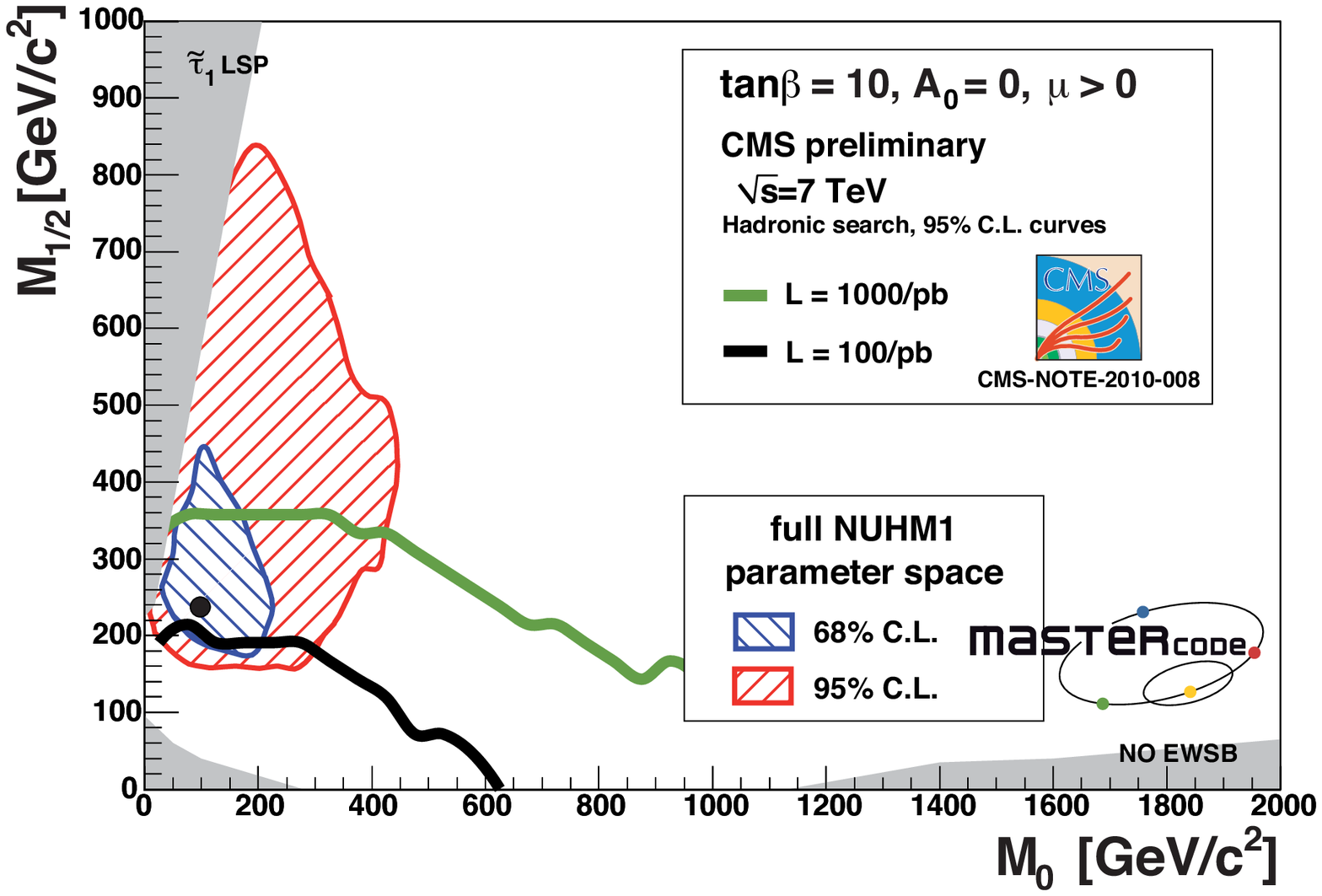}
\end{center}
\vspace{-2.5em}
\caption {The $(m_0, m_{1/2})$ plane in the CMSSM (left) and the
  NUHM1 (right).
  The dark shaded area at low $m_0$ and high $m_{1/2}$ is
  excluded due 
  to a scalar tau LSP, the light shaded areas at low $m_{1/2}$ do not
  exhibit electroweak symmetry breaking. 
  Shown in both plots are the best-fit point, indicated by a filled
  circle, and the 
  68 (95)\%~C.L.\ contours from our fit as dark grey/blue (light
  grey/red) overlays, scanned over all $\tan\beta$ and $A_0$ values.
  The 95\% C.L.\ exclusion curves (hadronic search channel) 
  at CMS with 1~(0.1)~fb$^{-1}$ at 7~TeV center-of-mass energy 
  is shown as green/light gray (black) solid curve.
} 
\label{fig:contours}
\end{figure*}

Our statistical treatment of the CMSSM and NUHM1 makes use of  
a large sample of points (about $2.5 \times 10^7$) 
in the SUSY parameter spaces 
obtained with the Markov Chain Monte Carlo
(MCMC) technique.
Our analysis is entirely frequentist, and avoids any
ambiguity associated with the choices of Bayesian priors.


\section{PREDICTIONS FOR SPARTICLE MASSES}

For the parameters of the best-fit CMSSM point we find
$m_0 = 60 \gev$,  $m_{1/2} = 310 \gev$,  $A_0 = 130 \gev$, $\tb = 11$
and $\mu = 400 \gev$, 
yielding the overall $\chi^2/{\rm N_{\rm dof}} = 20.6/19$ (36\% probability) 
and nominally $\Mh = 114.2 \gev$.
The corresponding parameters of
the best-fit NUHM1 point are $m_0 = 150 \gev$, $m_{1/2} = 270 \gev$,
$A_0 = -1300 \gev$, $\tb = 11$ and
$m_{h_1}^2  = m_{h_2}^2 = - 1.2 \times 10^6 \gev^2$ or, equivalently,
$\mu = 1140 \gev$, yielding
$\chi^2 = 18.4$ (corresponding to a similar fit probability as in the CMSSM)
and $\Mh = 120.7 \gev$. 

In Fig.~\ref{fig:contours} we display the best-fit value and the 
68\% and 95\% likelihood contours 
for the CMSSM (left plot) and the NUHM1 (right plot) in the 
$(m_0, m_{1/2})$ plane, obtained as described above from a fit taking into
account all experimental constraints.
We also show exclusion contours for the hadronic search mode (jets plus
missing energy) at CMS.
The green (light gray) solid line shows the 95\%~C.L.\ exclusion contour
for CMS for 1~\ifb\ at $\ecm = 7 \tev$~\cite{CMS7tev}. The black solid
line shows the corresponding results for only 0.1~\ifb.
(Similar results hold for ATLAS.)
One can see that with 1~\ifb\ the best-fit points can be tested,
together with a sizable part of the whole 68\%~C.L.\ preferred
regions. In the case of the NUHM1 (right plot) nearly the 68\%~C.L.\
region could be probed.

\begin{figure*}[htb!]
\centerline{
\includegraphics[height=11cm,width=7cm]{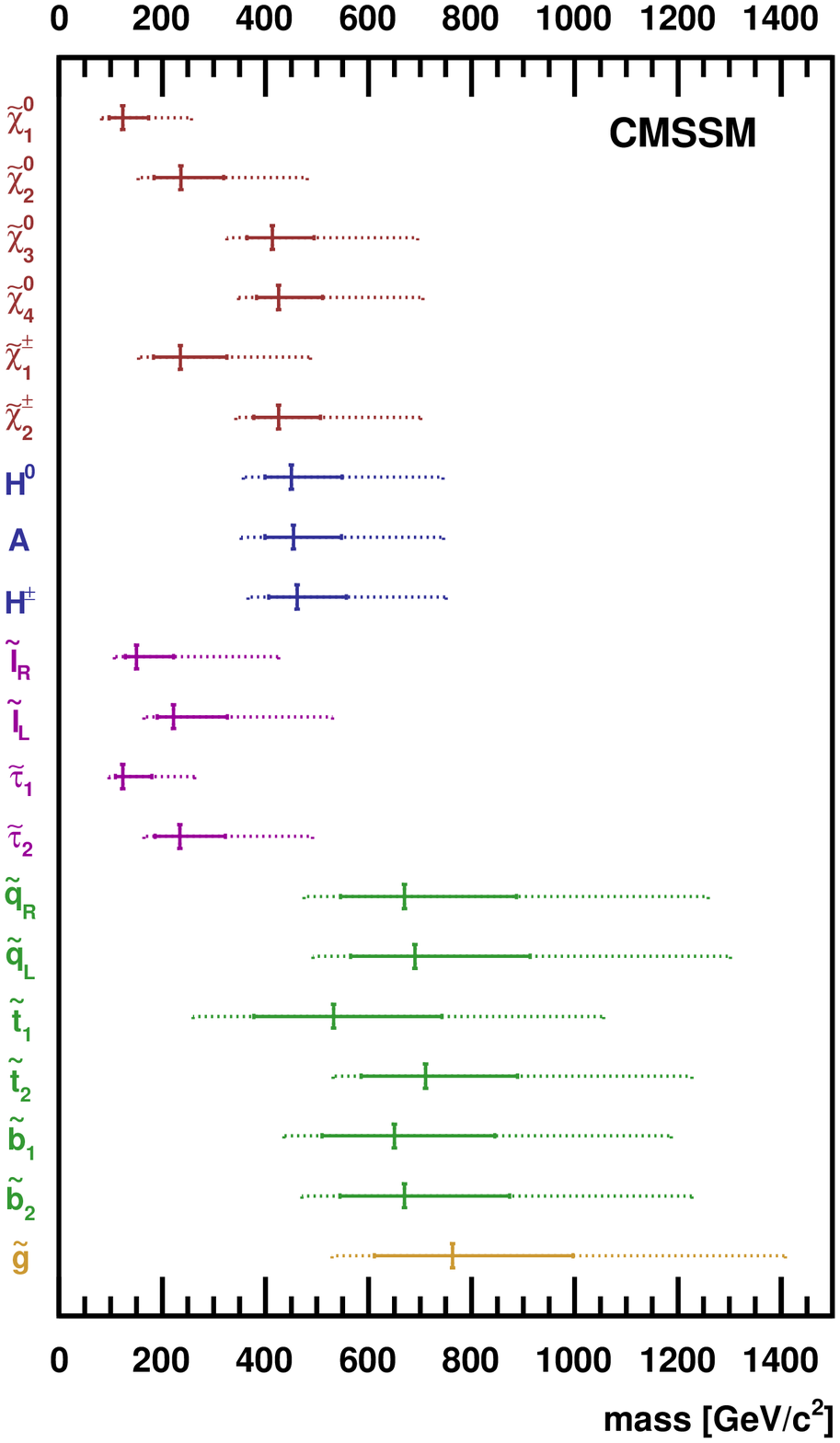}
\quad
\includegraphics[height=11cm,width=7cm]{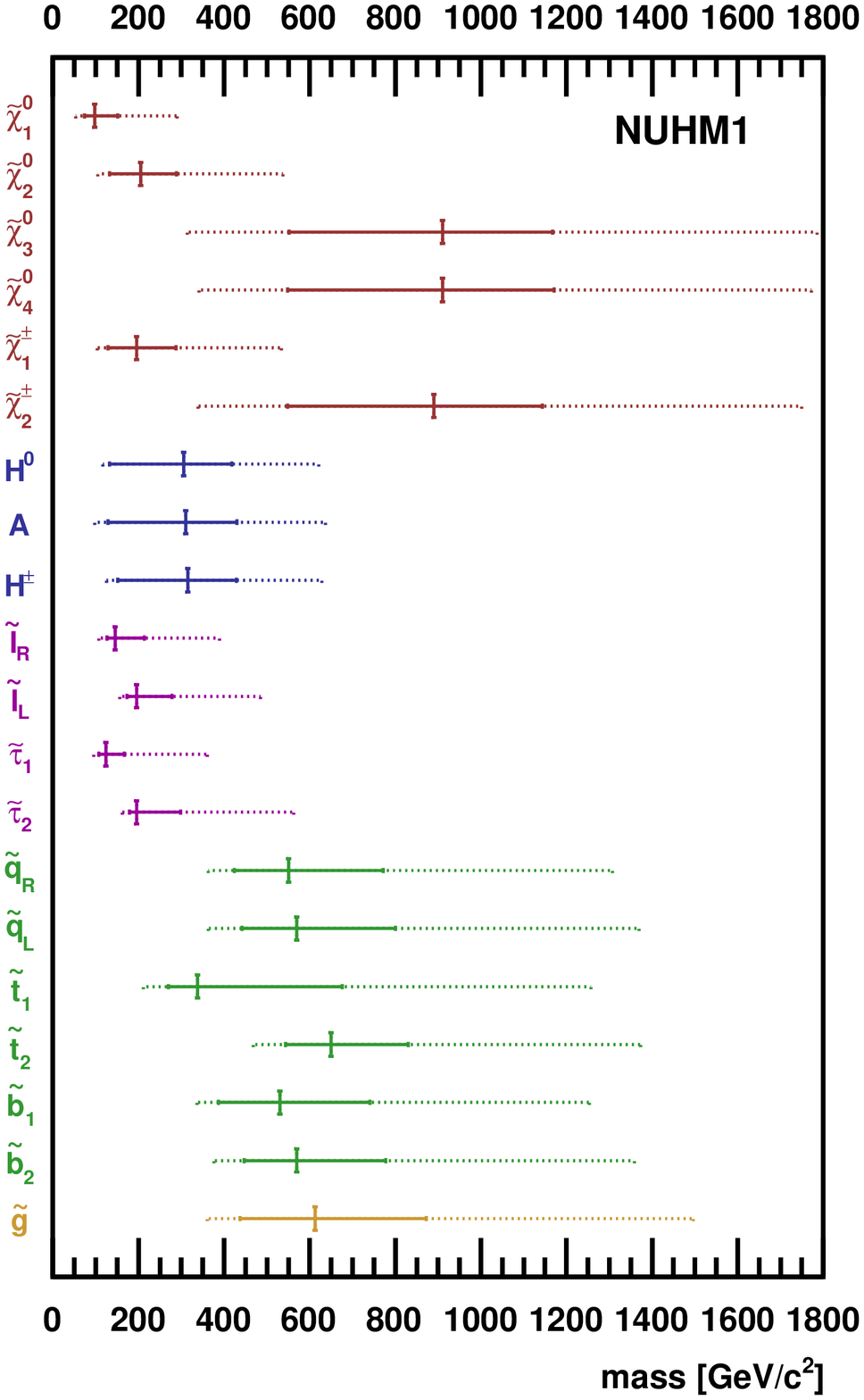}}
\vspace{-3em}
\caption{Spectra in the CMSSM (left) and the NUHM1
(right)~\cite{Master3}. The vertical 
solid lines indicate the best-fit values, the horizontal solid lines
are the 68\% C.L.\
ranges, and the horizontal dashed lines are the 95\% C.L.\ ranges for the
indicated mass parameters. 
}
\label{fig:spectrum}
\end{figure*}

We now review the results for the predictions of sparticles masses in the
CMSSM and the NUHM1, 
which are summarized in Fig.~\ref{fig:spectrum}. The results for the CMSSM
spectrum are shown in the left plot, and for the NUHM1 in the right plot. 
We start our discussion with the gluino mass, $\mgl$.
In both the CMSSM and the NUHM1, the
best-fit points have relatively low values of $\mgl \sim 750$ and $\sim
600 \gev$, respectively. These favored values are well within the range
even of the early operations of the LHC with reduced centre-of-mass
energy and limited luminosity, in accordance with the result
of \reffi{fig:contours}. However, even quite large values of  
$\mgl \lsim 2.5 \tev$
are allowed at the 3-$\sigma$ ($\Delta \chi^2 = 9$) level 
(not shown in Fig.~\ref{fig:spectrum}). The LHC
should be  able to discover a gluino with $\mgl \sim 2.5 \tev$ with
100/fb of integrated luminosity at 
$\ecm = 14 \tev$~\cite{atlastdr,cmstdr}, and the proposed 
SLHC luminosity upgrade to 1000/fb of integrated luminosity at 
$\ecm = 14 \tev$ should permit the discovery of a gluino with 
$\mgl \sim 3 \tev$~ \cite{Gianotti:2002xx}. 

The central values of the masses of the supersymmetric
partners of the $u, d, s, c, b$ quarks are slightly lighter than the
gluino, as seen in 
Fig.~\ref{fig:spectrum}. The difference between the gluino and the
squark masses is sensitive primarily to $m_0$.
The SUSY partners of the left-handed components of
the four lightest quarks, the ${\tilde q_L}$, 
are predicted to be slightly heavier than the corresponding right-handed
squarks, ${\tilde q_R}$, as seen by comparing the mass ranges in
Fig.~\ref{fig:spectrum}. As in the case of the gluino,
squark masses up to $\sim 2.5 \tev$ are allowed at the 3-$\sigma$
level. Comparing 
the left and right panels, we see that the squarks are predicted to be
somewhat lighter in the NUHM1 than in the CMSSM.

The scalar taus as well as the other scalar leptons are expected to be
relatively light, as can be seen in Fig.~\ref{fig:spectrum}. They would
partially be in the reach of the ILC(500) (i.e.\ with $\sqrt{s} = 500 \gev$)
and at the 95\% C.L.\ nearly all be in the reach of the
ILC(1000)~\cite{teslatdr,Brau:2007zza}. 
This also holds for the two lighter neutralinos and the light chargino.


\section{PREDICTIONS FOR THE LIGHTEST HIGGS MASS}

\begin{figure*}[htb!]
\centerline{\includegraphics[height=6.0cm]{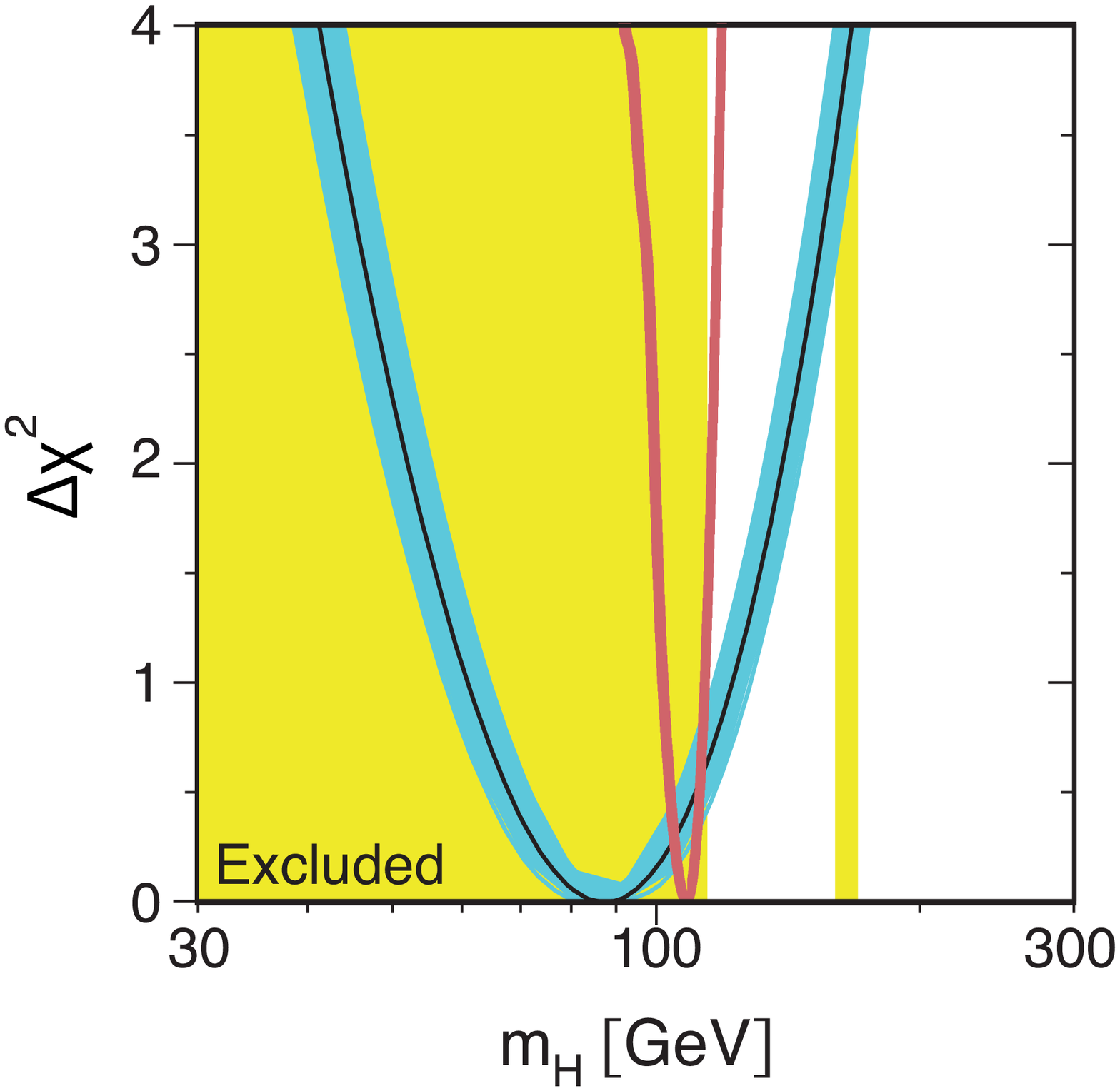}
\quad 
\includegraphics[height=6.0cm]{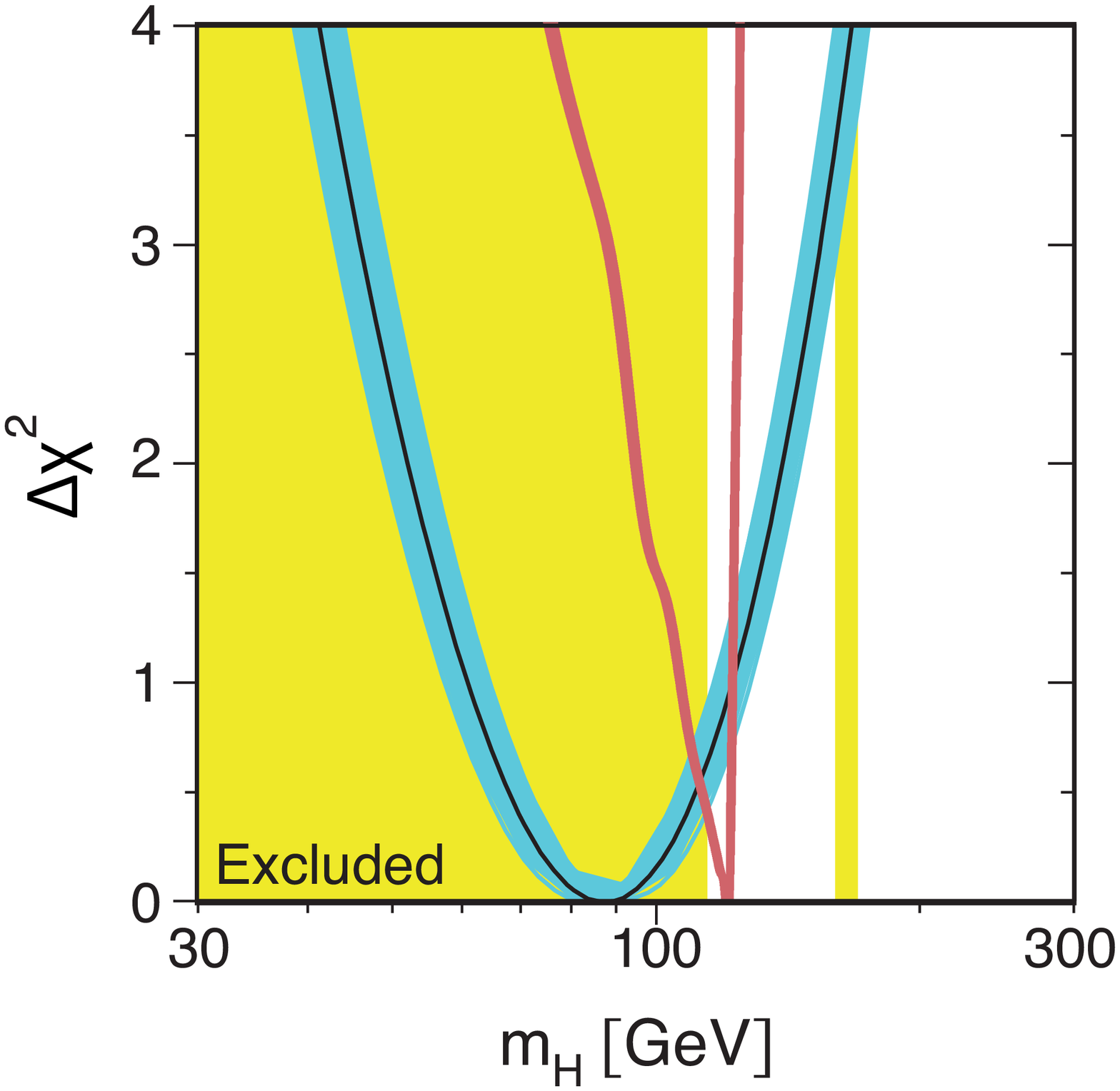}}
\vspace{-3em}
\caption{The $\chi^2$ functions for $\Mh$ in the CMSSM (left) and
  the NUHM1 (right)~\cite{Master3}, 
including the theoretical uncertainties (red bands). 
Overlayed is the blue band curve for the SM Higgs boson taken from
  \citere{lepewwg}. 
Also shown is the mass
range excluded for a SM-like Higgs boson (yellow shading).
The LEP limits are {\em not} included in the fit.
}
\label{fig:redband}
\end{figure*}

Finally we discuss 
the likelihood functions for $\Mh$ within the CMSSM and NUHM1 
frameworks obtained {\em when dropping} the
contribution to $\chi^2$ from the direct Higgs searches at LEP.
The results are 
shown in the left and right panels of Fig.~\ref{fig:redband},
respectively, together with the corresponding SM result~\cite{lepewwg}.

It is well
known that the central value of the Higgs mass in a SM
fit to the precision electroweak data lies below
100~GeV~\cite{lepewwg}, 
but the theoretical (blue bands in \reffi{fig:redband}) 
and experimental uncertainties 
in the SM fit are such that they are still compatible at the 
95\% C.L.\ 
with the direct lower limit of
114.4~GeV~\cite{Barate:2003sz} derived 
from searches at LEP. In the case of the CMSSM and NUHM1,
one may predict $\Mh$ on the basis of the underlying model
parameters, with a 1-$\sigma$ uncertainty of 1.5~GeV~\cite{Degrassi:2002fi},
shown as a red band in Fig.~\ref{fig:redband}. Also shown in
Fig.~\ref{fig:redband} are the LEP exclusion on a SM
Higgs~\cite{Barate:2003sz} 
and the Tevatron exclusion around $\MH \approx 165 \gev$~\cite{TevMH165} 
(yellow shading).
The LEP exclusion is directly applicable to the CMSSM,
since the $h$ couplings are essentially indistinguishable from
those of the SM Higgs boson~\cite{Ellis:2001qv,Ambrosanio:2001xb}, but
this is not necessarily the case in the NUHM1, as discussed earlier.

In the case of the CMSSM, we see in the left panel of
Fig.~\ref{fig:redband} that the minimum of the $\chi^2$
function occurs below the LEP exclusion limit. 
The fit result is still
compatible at the 95\% C.L.\ with the search limit, 
similarly to the SM case. As discussed above, 
a global fit including the LEP constraint has acceptable $\chi^2$.
In the case of the NUHM1, shown in the right panel of
Fig.~\ref{fig:redband}, we see that the minimum of the $\chi^2$
function occurs {\it above} the LEP lower limit on the mass of a SM 
Higgs at $\Mh \approx 121 \gev$. Thus, 
within the NUHM1 the combination of all other experimental
constraints {\em naturally} evades the LEP Higgs constraints, and no
tension between $\Mh$ and the experimental bounds exists.


\subsubsection*{Acknowledgements}

S.H.\ thanks the organizers of L\&L\,2010 for the invitation and the (as
always!) inspiring atmosphere.
We thank O.~Buchm\"uller, R.~Cavanaugh, A.~De~Roeck, J.~Ellis,
H.~Fl\"acher, G.~Isidori, K.~Olive and F.~Ronga with whom the results
presented here have been obtained.
Work supported in part by the European Community's Marie-Curie Research
Training Network under contract MRTN-CT-2006-035505
`Tools and Precision Calculations for Physics Discoveries at Colliders'
(HEPTOOLS).
